\documentclass{article}
\usepackage{amsmath}
\usepackage{graphicx}
\usepackage{psfrag}

\newcommand{\ket}[1]{\ensuremath{\left|#1\right>}}

\newcommand{\avrg}[1]{\ensuremath{\left<#1\right>}}

\newcommand{\etal}{\textit{et~al.}~}

\newcommand{\he}{$^4$He}
\newcommand{\const}{\text{const}\,}

\begin{document}
\title{Tunnelling defect nanoclusters in hcp \he\  crystals: alternative to supersolidity}
\author{A.F. Andreev\footnote{E-mail: andreev@kapitza.ras.ru}}
\date{}
\maketitle

\begin{center}
\textit{Kapitza Institute for Physical Problems, Russian Academy of Sciences,\\
Kosygin Str. 2, Moscow, 119334 Russia}
\end{center}

\begin{abstract}
A simple model based on the concept of resonant tunnelling
clusters of lattice defects is used to explain the low temperature
anomalies of hcp \he\  crystals (mass decoupling from a torsional
oscillator, shear modulus anomaly, dissipation peaks, heat
capacity peak). Mass decoupling is a result of an internal
Josephson effect: mass supercurrent inside phase coherent
tunnelling clusters. Quantitative results are in reasonable
agreement with experiments.
\end{abstract}


\psfragscanon

\section{Introduction}
Helium crystals are the most pronounced examples of the quantum
crystals in which quantum tunnelling of particles results in many
unusual phenomena. Delocalization of impurities, vacancies,
dislocations, and surface defects leads to quantum diffusion,
vacancy-induced mobility of impurities and ions, internal friction
anomalies, and weakly decaying crystallization waves, respectively
(see the review paper \cite{lit1}).

Motivated by theoretical predictions \cite{lit2,lit3,lit4} of the
superfluidity of solid \he, Kim and Chan (KC) performed
experiments \cite{lit5,lit6} similar to the Andronikashvili
experiment \cite{lit7,lit8} in which the decoupling of the
superfluid fraction of liquid HeII from a torsional oscillator
(TO) was discovered. The remarkable observation of KC was a
similar decoupling of a part of solid helium from TO below $0.2\,\text{K}$
which was interpreted as the superfluidity of a solid.

However, unlike the superfluid transition, the onset of the mass
decoupling is broad and is accompanied by a dissipation peak. Day
and Beamish \cite{lit9} measured the elastic shear modulus, and
observed a similar behavior (stiffening and dissipation) in the
same temperature range. Near the decoupling onset temperature, a
broad heat capacity peak was observed \cite{lit10}, but no
pressure-induced superflow through the solid was found
\cite{lit11,lit12}. The magnitudes of anomalies depend strongly on
the way the solid was prepared \cite{lit13}. All these anomalies
seem to be absent in perfect crystals.

In this paper (see also earlier letter \cite{lit14}), a simple
model is proposed to explain the low-temperature anomalies of
imperfect \he\ crystals. As in papers \cite{lit15,lit16}, the
concept of tunnelling two-level systems (TLSs) in solids
\cite{lit17} is used. In \cite{lit15,lit16} TLSs were considered
in highly disordered (glassy) samples. In this case, the
parameters of TLSs are uniformly distributed according to the
original tunnelling model \cite{lit18}. However, in the most
experiments, solid \he\ samples were grown by the blocked
capillary technique, but they consisted of poly-crystals. TLSs in
crystals are degenerate (or resonant) TLSs in which the bare (with
no tunnelling) energy difference of two localized states is zero
(see \cite{lit17} and below where a simple example is presented).
The physical reason of the degeneracy is the crystal symmetry. Two
localized states transform to each other under a crystal symmetry
transformation. (Otherwise, there is no reason for energy
difference to be small in crystals). The degenerate TLS is only
the simplest case. Tunnelling systems may generally consist of
more than two localized configurations \cite{lit17}.

In our simple model, we suppose that the main contribution to all
anomalies is introduced by degenerate TLSs of a certain (or
crystallographically equivalent) structure. In this case, the
characteristic parameters, including tunneling amplitudes, are
the same for all TLSs.

The key point is a peculiar quantum phenomenon of momentum deficit
for TLSs in moving solids \cite{lit15,lit16}. In a solid with
local velocity $\mathbf{v}$, the momentum of a TLS can under
certain conditions (see below) be equal to $m^*\mathbf{v}$, where
$m^*$ is the effective mass which is different from the
contribution $m$ of the TLS to the total mass. Generally, $m^*$
depends on the frequency and amplitude of the local velocity
oscillations, the difference $m^*-m$ being always negative.
Therefore, the TLS has nonzero internal momentum
$(m^*-m)\mathbf{v}$ directed opposite to the velocity of the
solid. In classical physics, a system of particles moving in a
restricted spatial region, has zero internal momentum up to the
frequencies on the order of the characteristic frequency of
particle oscillations. For TLSs, this frequency is the tunneling
frequency. We show below that under conditions of KC experiments,
TLSs in solid helium have the frequency independent internal
momentum down to the frequencies six orders of magnitude lower
than the tunneling frequency. The internal momentum disappears at
frequencies below the reciprocal transverse relaxation time (the
TLS phase memory time).

\section{TLS in imperfect crystals}
Simple example of a resonant TLS is a four-vacancy cluster in a
hcp \he\  crystal (see Fig.1). Vacancies are located in the apexes
of a tetrahedron. Three of them are disposed in the symmetry plane,
which is perpendicular to the c-axis of the crystal. The fourth
vacancy occupies one of the two positions which transform to each
other by the reflection in the plane. According to numerical
calculations \cite{lit19}, four-vacancy clusters in solid helium
are either unbound or are bound too weakly at the temperatures of
KC-experiments. However, highly metastable growth-introduced
vacancy clusters can persist in crystals for a long time. The same
can be true for clusters of a different metastable phase in
hcp-crystals \cite{lit20}. Large growth-introduced vacancy
clusters are probably the faceted liquid bubbles
studied experimentally in \he\ crystals \cite{lit21}.

\begin{figure}
\begin{center}
\includegraphics[scale=1.2]{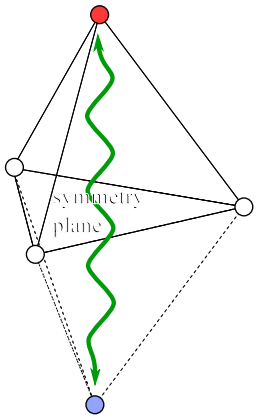}
\end{center}
\caption{Hypothetic arrangement of vacancies in a tetrahedronal cluster
with possible tunneling ``through'' the symmetry plane.} \label{tetra}
\end{figure}

The Hamiltonian of a TLS in a stationary crystal is $H_0 = \xi
\sigma_3 + \Delta \sigma_1$, where $\sigma_\alpha$, ($\alpha =
1,2,3$), are the Pauli matrices, $\Delta$ is the tunneling
amplitude, and $\mp \xi$ ($\xi>0$) are the bare energies of two
localized states. Here, the TLS is supposed to be nondegenerate,
because we consider below a shear deformation that breaks the
symmetry.

\section{TLS in rotating crystals}
The Hamiltonian of the TLS in a crystal moving with a local
velocity $\mathbf{v}$, according to Galilean transformation is
$H_1 = H_0 +\mathbf{p}\mathbf{v}$, where $\mathbf{p}$ is the TLS
momentum operator. We suppose that the TLS tunneling is
accompanied by the displacement of a mass $m$ by a vector
$\mathbf{a}$. The operator part of the TLS center of gravity
coordinate can be written as $\mathbf{r} = -\sigma_3\mathbf{a}/2$.
The momentum operator is $\mathbf{p} = m\dot{\mathbf{r}} =
(im/\hbar)[H_0 \mathbf{r}] = -(m\Delta/\hbar)\mathbf{a}\sigma_2$
\cite{lit15,lit16}.

The internal momentum in the TLS state $\psi_1 \ket{1} + \psi_2
\ket{2}$ is proportional to $\avrg{\sigma_2} \propto
\sin(\varphi_1 - \varphi_2)$. Here, \ket{1} and \ket{2} are the
localized states, and $\varphi_1$ and $\varphi_2$ are the phases of
$\psi_1$ and $\psi_2$, respectively. Therefore, the internal momentum is
caused by a kind of internal Josephson effect: the mass
supercurrent inside phase coherent TLSs.

We suppose that the velocity is a result of an axisymmetric
container rotation. Otherwise, additional terms should be added to
the Hamiltonian to take the macroscopic displacement of the
container walls into account (see \cite{lit22}, \S 11). Because
the size of the TLS is assumed to be much smaller than the length
scale of the rotating container, we can use the following
expressions for the velocity and the TLS angular momentum:
$\mathbf{v} = \mathbf{\Omega}\times\mathbf{R}$, and $M
=(\mathbf{R}\times\mathbf{p})_z$, where $\mathbf{\Omega}$ is the
angular velocity and $\mathbf{R}$ is the
coordinate of the TLS ($\mathbf{R}\perp\mathbf{\Omega}$)  with respect to the origin located at the
rotation axis which is parallel to the $z$-axis. We have
\begin{equation}
H_1 = H_0 + M\Omega, \label{eq1}
\end{equation}
where $M = \mu\sigma_2$,
$\mu = -(m\Delta/\hbar)aR\cos\theta$,
$\theta$ is the angle between
$\mathbf{a}$ and $\mathbf{v}$.

After canonical transformation by the unitary operator
\begin{equation}
U = (2\varepsilon(\varepsilon+\xi))^{-1/2}
(\varepsilon+\xi-i\Delta\sigma_2), \label{eq2}
\end{equation}
where $\varepsilon=(\Delta^2 + \xi^2)^{1/2}$, we obtain the
Hamiltonian in the more convenient form
\begin{equation}
H = UH_1U^+ = -\varepsilon\sigma_3 + \mu\Omega\sigma_2.
\label{eq3}
\end{equation}
This Hamiltonian can be written in the form
\begin{equation}
H = -h_\alpha \sigma_\alpha, \label{eq4}
\end{equation}
where $h_\alpha$ is the ``field'' with components $h_1=0$,
$h_2=-\mu\Omega$, and $h_3=\varepsilon$.

The TLS density matrix $w$ is generally determined by the real
polarization vector $s_\alpha$:
\begin{equation}
w = (1 +s_\alpha \sigma_\alpha)/2. \label{eq5}
\end{equation}
We have
\begin{equation}
\avrg{\sigma_\alpha} = \mathrm{Tr}\,(w\sigma_\alpha) = s_\alpha.
\label{eq6}
\end{equation}
From the equation for the density matrix
\begin{equation}
\dot{w} = (i/\hbar)[w,H], \label{eq7}
\end{equation}
we obtain the dynamic equation for a free TLS (without
dissipation):
\begin{equation}
\hbar\dot{s}_\alpha = e_{\alpha \beta \gamma}h_\beta
s_\gamma,\label{eq8}
\end{equation}
where $e_{\alpha \beta \gamma}$ is the Levi-Civita tensor.

The adiabatic theorem (see \cite{lit23}, chap II, \S 5c) takes
place as a consequence of (8): along with the modulus $s = |s_\alpha|$
of polarization, the angle between the field $h_\alpha$ and
$s_\alpha$ is the integral of motion. The process is adiabatic if
the time scale of the field variation is much longer than
$\hbar/|h_\alpha|$. The last condition is fulfilled with a
significant safety margin: the characteristic frequency of the field
variation in experiments mentioned above, is on the order of
$1\,\text{kHz}$, but the magnitude of the field $|h_\alpha|$ is on 
the order of $\Delta \sim 0.1\,\text{K}$.

Until the rotation (and deformation) is applied, the polarization
is directed along the field, and the absolute value of the
equilibrium polarization is $s=s_0$, where $s_0 =
\tanh\varepsilon/T$. According to the adiabatic theorem, we have
for the free TLS:
\begin{equation}
s_1 = 0, \quad
s_2 = -\frac{\mu}{E}\Omega s, \quad
s_3 = \frac{\varepsilon}{E} s = s -
\frac{\mu^2\Omega^2}{E(E+\varepsilon)}s,
\label{eq9}
\end{equation}
where $E=\left(\varepsilon^2 + \mu^2\Omega^2\right)^{1/2}$ is
the field modulus.

We suppose that equations describing relaxation
processes in a stationary solid have the standard form:
\begin{equation}
\dot{s_2} = -\frac{s_2}{\tau_2},\quad
\dot{s_3} = -\frac{s_3 -s_0}{\tau_1},
\label{eq10}
\end{equation}
where $\tau_1$ and $\tau_2$ are the longitudinal and transverse
relaxation times, respectively. Additional terms to the time
derivatives (10) taking the rotation of the solid into
account, can be determined from (9). Finally, we obtain the following
dynamic equations for the TLS:
\begin{equation}
\frac{\partial}{\partial t}
\left(s_2 +
\frac{\mu\Omega}{E}s\right) = -\frac{s_2}{\tau_2},\quad
\frac{\partial}{\partial t}\left(s_3 + \frac{\mu^2
\Omega^2}{E(E+\varepsilon)}s\right) = -\frac{s_3-s_0}{\tau_1}.
\label{eq11}
\end{equation}
Following the work of Burin \etal\cite{lit24}, we will assume that
both reciprocal relaxation times depend linearly on temperature:
\begin{equation}
\tau_1^{-1} = \chi_1 T,\quad  \tau_2^{-1} = \chi_2 T,
\label{eq12}
\end{equation}
where $\chi_1$ and $\chi_2$ are constants.

\section{Torsional oscillations}
Consider torsional oscillations with a small amplitude
$\Omega(t)\propto\exp(-i\omega t)$. From the first of the
equations (11), we obtain the mean value of the TLS angular
momentum 
$\avrg{M} = \mu s_2 = I(\omega)\Omega$, where
\begin{equation}
I(\omega) = \frac{\mu^2}{\Delta} \frac{i\omega\tau_2 - \omega^2
\tau_2^2}{\omega^2 \tau_2^2 + 1} \tanh\frac{\Delta}{T}
 \label{eq13}
\end{equation}
is the TLS rotational inertia. The real part $I'(\omega)$ of $I$
determines the TLS mass deficit $\delta m(\omega) = m - m^* =
-I'(\omega)/R^2$. The crystal fraction $f$, decoupled from TO, is
 $f = n\delta m/\rho$, where $\rho$ is the crystal density,
and $n$ is the number of  resonant TLSs per unit volume. We have
\begin{equation}
\delta m(\omega) = \left(\frac{ma\cos\theta}{\hbar}\right)^2
\frac{\omega^2 \tau_2^2}{\omega^2 \tau_2^2 + 1}
\Delta\tanh\frac{\Delta}{T}. \label{eq14}
\end{equation}
The mass deficit of a free TLS ($\omega\tau_2 \gg 1$) is frequency
independent and coincides with the result of \cite{lit16}.

For uniform ($\omega = 0$) rotation $\delta m = 0$, in accordance
with \cite{lit15,lit16}. The intermediate ``quasi-equilibrium''
region which was considered in \cite{lit15,lit16}, is absent in
our simple model.

Korshunov \cite{lit25} calculated the TLS mass deficit for
translational oscillations of a closed container filled with solid
helium. His value of $\delta m$ is proportional to the extremely
small factor $(\omega/\Delta)^2$. In fact, the Korshunov result is
an explanation of the absence of TO anomalies in the blocked
annulus experiments \cite{lit6,lit26}.

\begin{figure}
\begin{center}
\includegraphics[scale=.7]{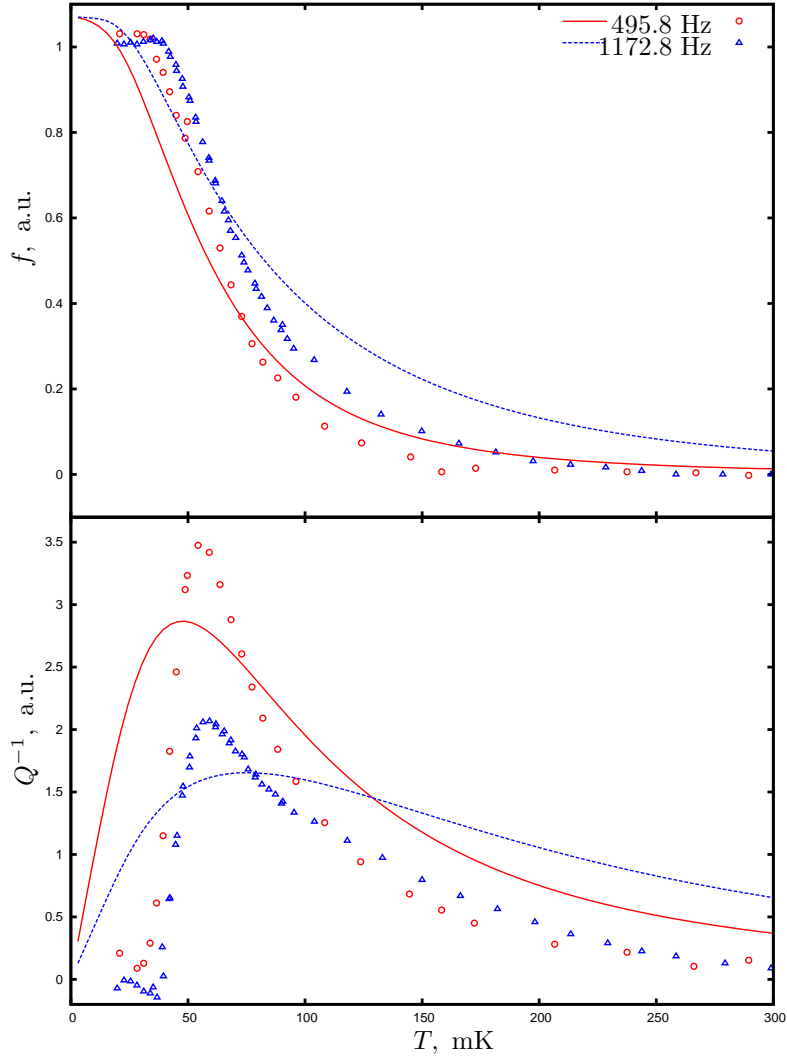}
\end{center}
\caption{Temperature dependencies of decoupled fraction~\eqref{eq16}
and dissipation~\eqref{eq15} ($\Delta=51\,\text{mK}$,
$\chi_2=5.9\,\text{kHz}/\text{K}$) for two frequencies
fitting experimental data~\cite{lit27}.} \label{ncri}
\end{figure}

The contribution $\avrg{\smash{\dot{E}}} = \avrg{\Omega\smash{\dot{M}}}$ of the TLS to
the energy dissipation is determined by the imaginary part
$I''(\omega)$ of $I$ (see \cite{lit22}, \S 123): $\avrg{\smash{\dot{E}}\vphantom{E}} =
(\omega/2)I''(\omega)\left|\Omega\right|^2$.
Under experimental conditions \cite{lit27}, the dependence of the reciprocal
quality factor of TO on temperature and frequency, is determined
by the expression
\begin{equation}
Q^{-1} \propto  \frac{\omega \tau_2}{\omega^2 \tau_2^2 +
1}\tanh\frac{\Delta}{T}. \label{eq15}
\end{equation}
The temperature dependence of (15) for two different frequencies
$\omega$ are shown at the bottom of Fig.2. Note that according to
(15), these curves intersect at some point.

The corresponding formula for the decoupled fraction $f$,
according to (14), is
\begin{equation}
f \propto  \frac{\omega^2 \tau_2^2}{\omega^2 \tau_2^2 +
1}\tanh\frac{\Delta}{T}. \label{eq16}
\end{equation}
The temperature dependencies of (16) for the same two frequencies
are shown at the top of Fig.2.

\section{Nonlinear torsional oscillations}
Consider nonlinear torsional oscillations at low temperatures
(free TLSs). The oscillations are described by the equation
\begin{equation}
\ddot{\alpha} + \omega_0^2\alpha = -(1/I_0)\dot{M}(t)
\label{eq17}
\end{equation}
for the rotation angle $\alpha$ ($\dot{\alpha}=\Omega$). Here,
$I_0$ and $\omega_0$ are the rotational inertia and the TO
fundamental frequency in the absence of TLSs, respectively. $M(t)$
is the TLS angular momentum. According to (1) and (9) we have
\begin{equation}
M(t) = -\frac{R\Delta}{v_c^2}
        \frac{v\cos^2\theta}{(1+\cos^2\theta(v/v_c)^2)^{1/2}},
\label{eq18}
\end{equation}
where $v_c = \hbar/(ma)$ is a characteristic (critical) velocity.

In the absence of TLSs, the oscillations are harmonic:
$\alpha(t)=\alpha_0 \cos\varphi$, where $\alpha_0$ is the
amplitude and $\varphi=\omega_0 t$. The time dependence of $M$ is
determined by (18) with $v=v_0 \sin\varphi$, $v_0 = -\alpha_0
\omega_0 R$. Being the periodic function of $\varphi$, $M(t)$ can
be represented by its Fourier series. It is known \cite{LL1} that a small
correction to the fundamental frequency is determined by the
resonant part (proportional to $\cos\varphi$) of
the right hand side of \eqref{eq17}.
Resonant contribution to the momentum $M_\text{res}(\varphi)$ is proportional to
$\sin\varphi$
\begin{equation}
M_\text{res}(\varphi) = \sin\varphi \frac{1}{\pi}\int_{-\pi}^{\pi}
M(\varphi')\sin\varphi'd\varphi'.
\label{eq19}
\end{equation}
According to (18) and (19), the TLS contribution to the
fundamental frequency and, therefore, to the value of the
decoupled fraction $f(v_0)$ is determined by the expression
\begin{equation}
f(v_0) = \const \cos^2 \theta \int_{-\pi}^{\pi}d\varphi
\frac{\sin^2\varphi}{(1+\cos^2\theta (v_0/v_c)^2
\sin^2\varphi)^{1/2}}.
\label{eq20}
\end{equation}
We consider a narrow annulus where $R\approx \const$ for all
TLSs. For poly-crystals, the expression (20) should be averaged
over all directions of $\mathbf{a}$ at a given $\mathbf{v}$. We
have
\begin{equation}
f(v) = \const \int_{-1}^{1}\cos^2 \theta d\cos\theta
\int_{-\pi}^{\pi} \frac{\sin^2\varphi
d\varphi}{(1+(v/v_c)^2\cos^2\theta \sin^2\varphi)^{1/2}}.
\label{eq21}
\end{equation}
Here and below we omit the index $0$ in the velocity amplitude
$v_0$. Finally, we obtain
\begin{equation}
\frac{f(v)}{f(0)} = \frac{12}{\pi} \int_{0}^{1}x^2 dx
\int_{0}^{\pi/2} \frac{\sin^2\varphi d\varphi}{(1+(v/v_c)^2 x^2
\sin^2\varphi)^{1/2}}. \label{eq22}
\end{equation}
This dependence is plotted in Fig.3. According to (14), the mass
deficit $\delta m$ for a single TLS is on the order of
$\Delta/v_c^2$. Assuming $\Delta\sim 0.1\,\text{K}$ and $v_c\sim 10\,\mu\text{m}/\text{s}$,
we obtain $\delta m\sim 10^{-11}\,\text{g}$. The experimental value of the
decoupled fraction $f=n\delta m/\rho \sim 10^{-3}$ corresponds to
the quite low TLS concentration $n\sim 10^7\,\text{cm}^{-3}$. The
characteristic size $L$ of the defect clusters can be estimated
from the value of the critical velocity. Assuming that $ma=\rho
L^4$, we have $L\sim 10\,\text{nm}$.

\begin{figure}
\begin{center}
\psfrag{present}[l][l]{present model}
\psfrag{xaxis}[c][c]{$v,\ \mu\text{m}/\text{s}$}
\psfrag{yaxis}[l][l]{$f(v)/f(0)$}
\includegraphics[scale=.8, bb = 0 0 365 270]{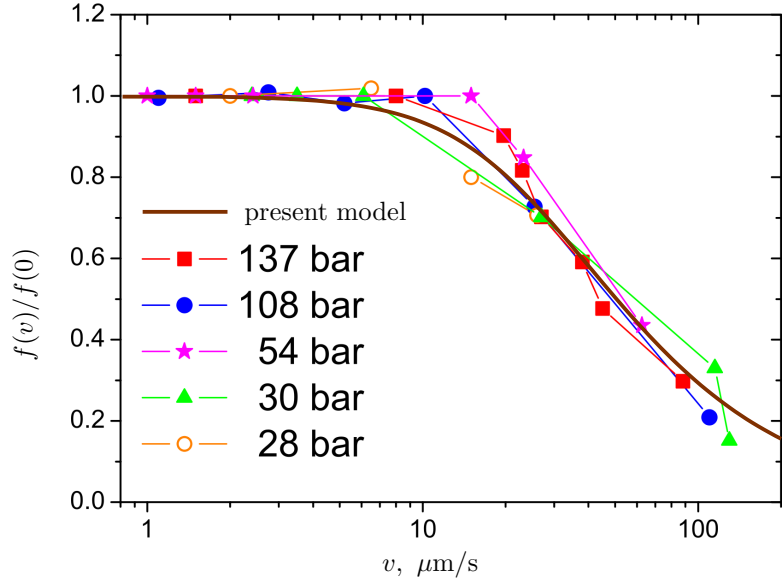}
\end{center}
\caption{Normalized decoupled fraction~\eqref{eq22}
($v_c=17.25\,\mu\text{m}/\text{s}$)
versus oscillation velocity amplitude
fitting experimental data~\cite{lit29}.} \label{crit}
\end{figure}

\section{The shear modulus anomaly}
Let $\zeta = \partial u_x/\partial y$ be a shear strain ($u_x$ is
the displacement along the $x$-axis). According to (3) in the
absence of rotation the derivative of the TLS Hamiltonian with
respect to the strain is $\partial H/\partial \zeta = -\sigma_3
\partial \varepsilon/\partial\zeta$. The TLS contribution to the
stress $\sigma$ is determined by the mean value of this derivative
\begin{equation}
\sigma = n<\partial H/\partial \zeta> = -ns_3
\partial\varepsilon/\partial\zeta. \label{eq23}
\end{equation}

\begin{figure}
\begin{center}
\includegraphics[scale=.5]{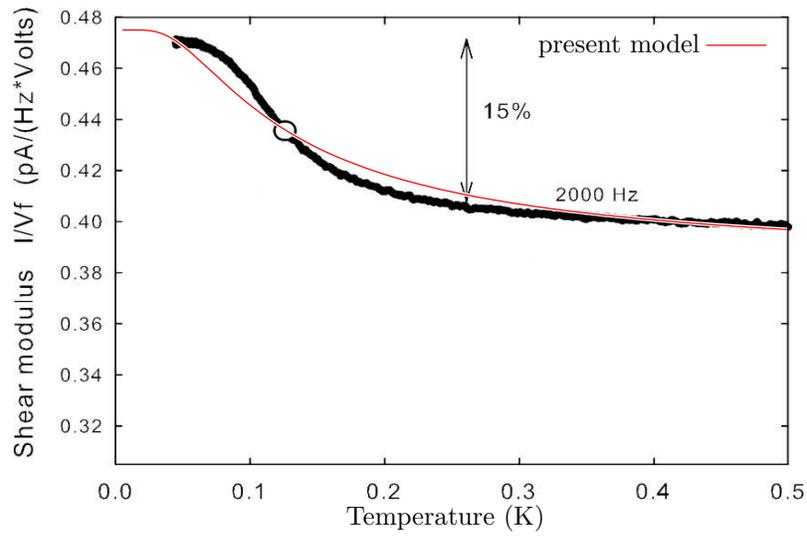}
\end{center}
\caption{Experimentally observed \cite{shear-lit}
temperature dependence of shear modulus at high frequency and a fit for
$\Delta=0.08\,\text{K}$.} \label{shear}
\end{figure}

We consider small oscillations of the strain $\zeta(t)
\propto\exp(-i\omega t)$. The oscillating part of the TLS
polarization $s_3'$ and that of the instantaneous equilibrium
polarization $s_0'$ satisfy, according to (10), the equation
$i\omega s_3' = (s_3' - s_0')/\tau_1$, where
\begin{equation}
s_0' = \zeta \frac{\partial}{\partial \zeta}
\tanh\frac{\varepsilon}{T}  =\frac{\zeta}{T\cosh^2(\Delta/T)}
\left(\frac{\partial \varepsilon}{\partial\zeta}\right)_0.
\label{eq24}
\end{equation}
Here, $(\partial\varepsilon/\partial\zeta)_0$ is the derivative
$\partial\varepsilon/\partial\zeta$ at $\zeta = 0$. We have
\begin{equation}
s_3' =
\frac{1}{1-i\omega\tau_1}\frac{\zeta}{T\cosh^2(\Delta/T)}\left(\frac{\partial
\varepsilon}{\partial\zeta}\right)_0. \label{eq25}
\end{equation}
The oscillating part of the stress is
\begin{equation}
\sigma' = -n\left\{ \zeta\left(\frac{\partial^2
\varepsilon}{\partial\zeta^2}\right)_0 \tanh\frac{\Delta}{T} +
\left(\frac{\partial\varepsilon}{\partial \zeta}\right)_0
s'_3\right\}. \label{eq26}
\end{equation}
The TLS contribution to the complex shear modulus $G(\omega)$
defined by the expression $\sigma' = G(\omega) \zeta$, is
\begin{equation}
G(\omega) = n\left\{ \left| \frac{\partial^2
\varepsilon}{\partial\zeta^2}\right| \tanh\frac{\Delta}{T}
-\frac{1}{1-i\omega\tau_1}\frac{1}{T\cosh^2(\Delta/T)}\left(\frac{\partial
\varepsilon}{\partial\zeta}\right)^2_0 \right\}. \label{eq27}
\end{equation}
The second derivative $\partial^2 \varepsilon / \partial \zeta^2$
should be negative. Otherwise, the real part of $G$, contrary to
experiments \cite{lit9}, would be negative.

The dissipation of the shear oscillations is determined by the
imaginary part of $G$. Its temperature dependence is characterized
by the presence of a characteristic peak, in accordance with
experiments \cite{lit9}.

It was noted \cite{lit9} that $G'$ and $f$ change very similarly
with temperature. This is consistent with our results: due to (16)
and (27), in high frequency region $\omega\tau \gg 1$, we have $G
\propto f \propto \tanh(\Delta/T)$. This dependence for $G$ is
plotted in Fig.4.

\begin{figure}
\begin{center}
\includegraphics[scale=.8]{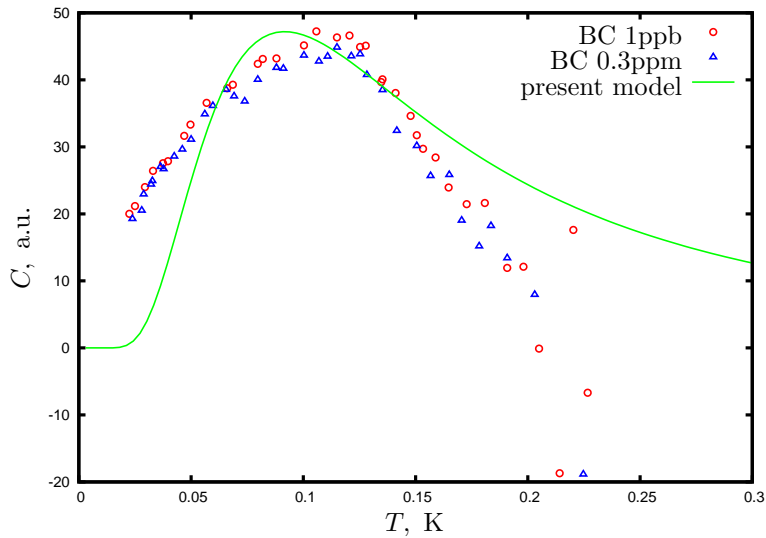}
\end{center}
\caption{Temperature dependence of heat capacity
\eqref{eq29} ($\Delta=0.11\,\text{K}$) and experimental
data~\cite{lit10}.} \label{heatcap}
\end{figure}

\section{Thermodynamics of resonant TLSs}
The contribution of resonant TLSs to the free energy of the unit
volume of a crystal is
\begin{equation}
F = -nT\log\left(2\cosh\frac{\Delta}{T}\right).
\label{eq28}
\end{equation}
The heat capacity is determined by the well-known formula:
\begin{equation}
C = -T\frac{\partial^2 F}{\partial T^2} = n\left(\frac{\Delta}{T}\right)^2
\cosh^{-2} \frac{\Delta}{T}. \label{eq29}
\end{equation}
This dependence is plotted in Fig.5.

The contribution of TLSs to pressure is
\begin{equation}
P = \rho\left(\frac{\partial F}{\partial \rho}\right)_T =
-n\rho\frac{\partial\Delta}{\partial\rho} \tanh\frac{\Delta}{T}.
\label{eq30}
\end{equation}
The $P(T)$ dependence at a fixed volume was measured by Grigoriev
at al. \cite{lit28}. The glassy contribution ($\propto T^2$) was
found. The term (30) was absent. The possible reason is that the
derivative $\partial \Delta/\partial \rho$ is anomalously small.
Indeed, this derivative determines in our model the pressure
dependence of the onset temperature of the mass decoupling. KC
\cite{lit29} observed no apparent change of the onset temperature
with pressure.

\section{Summary}
Low-temperature anomalies of hcp \he\ crystals are caused by
growth-introduced resonant nanoscale clusters of lattice defects.
The quantum phase coherence and tunnelling of the clusters
are crucial. The important problem is to identify the structure
of the clusters. The most probable candidates seem to be the
faceted liquid bubbles.

\section{Acknowledgments}
I thank L.A.Melnikovsky for helpful discussions. This work was
supported by the Russian Foundation for Basic Research, project
no. 09-02-00567 and by grant NSh-65248.2010.2 under the Program
for Support of Leading Science Schools.

\end {document}